\documentclass[aps,prd,preprint,a4paper,showpacs,nofootinbib,superscriptaddress]{revtex4-2}
\usepackage{bm}
\usepackage{indentfirst}
\usepackage{amsmath}
\usepackage{graphicx}
\usepackage{amssymb}
\usepackage{subfigure}
\usepackage{amssymb}
\usepackage{hyperref}
\usepackage{epstopdf}
\usepackage{cancel}
\usepackage[section]{placeins}

\newcommand{\sech}{\operatorname{sech}}

\usepackage[utf8]{inputenc}
\hypersetup{
    colorlinks=true,
    linkcolor=red,
    citecolor=blue,
}
\usepackage{color}
\usepackage[T1]{fontenc}
\usepackage{txfonts}
\usepackage{orcidlink}
\usepackage[title]{appendix}	

\begin{document}

\title{A Critical Point on the Hairy Black Hole Phase Boundary in the Improved Holographic Einstein-Maxwell-Dilaton Theory}

\author{Hong Guo}
\email{guohong@ibs.re.kr}
\affiliation{Escola de Engenharia de Lorena, Universidade de São Paulo, 12602-810, Lorena, SP, Brazil}

\author{Wei-Liang Qian}
\email{wlqian@usp.br}
\affiliation{Escola de Engenharia de Lorena, Universidade de São Paulo, 12602-810, Lorena, SP, Brazil}
\affiliation{Faculdade de Engenharia de Guaratinguet\'a, Universidade Estadual Paulista, 12516-410, Guaratinguet\'a, SP, Brazil}
\affiliation{Center for Gravitation and Cosmology, College of Physical Science and Technology, Yangzhou University, 225009, Yangzhou, China}

\author{Bean Wang}
\affiliation{Department of Physical Sciences and Applied Mathematics, Vanguard University, Costa Mesa, CA 92626, USA}

\begin{abstract}
In this work, we investigate the hairy black hole solutions and their dual phase diagram in the improved holographic Einstein–Maxwell–Dilaton (EMD) model.
From the gravitational perspective, the rich phase structures observed in the dual boundary field theory originate from the intricate interplay between the scalar field formalism and the Maxwell field coupling mechanism.
Two distinct types of hairy black hole solutions are found in this framework.
Type-I hairy black holes are predominantly governed by scalar potential dynamics, whereas Type-II solutions emerge through nonminimal coupling to the $U(1)$ gauge field.
We map out the phase distribution in the $(\mu_B,T)$ parameter plane and delineate the boundary separating these two hairy phases.
The phase diagram exhibits a first-order phase transition line consistent with previous findings, accompanied by a subtle third-order phase transition line that terminates at a critical point positioned at the turning point of the entire phase boundary curve.
Our results complement existing research on holographic EMD theory by offering a comprehensive characterization of phase distributions, transition boundaries, and their gravitational sector interpretations.
These insights will enable more effective engineering of specific phase structures for simulating strongly coupled systems through targeted modifications to the EMD model.
\end{abstract}

\maketitle

\newpage

\section{Introduction}\label{sec=intro}

The Beam Energy Scan (BES) program at Brookhaven's Relativistic Heavy Ion Collider (RHIC) stands at the forefront of probing the most elusive aspects of finite-density Quantum Chromodynamics (QCD)~\cite{Bzdak:2019pkr,Busza:2018rrf,Lovato:2022vgq}.
As the fundamental theory of strong interactions, QCD provides a theoretical framework for investigating phase transitions of nuclear matter under extreme conditions, particularly the transition from confined hadronic matter to a deconfined quark-gluon plasma (QGP). 
The exploration of this transition, and more broadly the study of the QCD phase diagram, has long been a central pursuit in high-energy physics and continues to be one of its core research topics~\cite{Fu:2022gou,Guenther:2020jwe,Du:2024wjm,Luo:2020pef}.
These forefront studies not only deepen our understanding of the early universe evolution and the formation of matter, revealing non-perturbative effects of strong interactions and the complex dynamical behavior of many-body systems, but also provide crucial insights into cosmology, gravitational waves, and potential new physics~\cite{Espinosa:2010hh,Boeckel:2011yj,Caprini:2015zlo,Pasechnik:2023hwv,Liu:2021svg,Shao:2022oqw}.

This transition, occurring at sufficiently high temperature or baryon density, is one of the fundamental predictions of QCD.
At negligible baryon chemical potential, the transition exhibits a crossover behavior, but under extreme conditions, various theoretical models suggest that the transition may become a first-order phase transition once the chemical potential exceeds a critical value.
Therefore, the existence of a critical endpoint (CEP) is a defining feature of the QCD phase diagram, marking both the termination point of the first-order phase transition and the onset of crossover transitions~\cite{Stephanov:1998dy}. 
Although the first phase of BES experiments probed a limited baryon chemical potential range $\mu_B/T \lesssim 3$~\cite{STAR:2017sal}, the wealth of experimental data accumulated has already revealed the crossover nature of the transition from hadronic matter to the QGP~\cite{Du:2024wjm}.
The ongoing second phase of BES, together with forthcoming heavy-ion programs at FAIR~\cite{Lovato:2022vgq,Senger:2022bjo}, NICA~\cite{Senger:2022bjo}, and J-PARC~\cite{Aoki:2021cqa}, will extend research into higher density regions, significantly enhancing our capability to locate the CEP and map the QCD phase diagram.

Current theoretical investigations of the QCD phase diagram primarily rely on first-principles lattice QCD calculations.
At finite temperatures with zero or small baryon chemical potential, the equation of state can be reliably determined, enabling precise simulations of the crossover transition between hadronic matter and the QGP.
However, the inherent sign problem~\cite{Allton:2002zi,Allton:2005gk,Bazavov:2017dus,Borsanyi:2021sxv} in lattice simulations prevents the convergence at high baryon densities, thereby limiting their applicability to the study of the QCD phase diagram under extreme conditions.
A natural alternative is to extrapolate the lattice QCD data, obtained at zero or small baryon chemical potentials and temperatures around the crossover transition temperature, to infer and extend the phase structure into the higher density regime, with particular focus on identifying the location of CEP marking the termination of first-order deconfinement phase transitions.
Several theoretical frameworks have been extensively employed in analyzing the QCD phase diagram, including the functional renormalization group (FRG)~\cite{Fu:2022gou}, Dyson–Schwinger equations (DSEs)~\cite{Roberts:1994dr,Qin:2010nq,Shi:2014zpa,Gao:2016qkh}, random matrix models (RMM)~\cite{Halasz:1998qr}, and the Nambu–Jona-Lasinio (NJL) model~\cite{Hatsuda:1994pi,Schwarz:1999dj}, among others.
It is important to note that existing experimental data have yet to reveal any sign of a CEP in the range $\mu_B/T \lesssim 3$~\cite{STAR:2017sal}, where various lattice QCD models can still be reliably extended~\cite{deForcrand:2009zkb,Ding:2015ona}.
Current attempts remain inconclusive regarding the existence of the CEP, including the deconfinement phase transition.
The QCD phase diagram under extreme conditions thus contains largely uncharted territory, warranting further investigation both theoretically and experimentally.

The gauge/gravity duality~\cite{Maldacena:1997re,Gubser:1998bc,Witten:1998qj} has attracted considerable attention over the past decades, particularly following the consistency between RHIC experiments on the viscosity/entropy-density ratio and holographic predictions~\cite{Kovtun:2003wp,Buchel:2003tz,Ge:2008ni}.
Due to the intrinsic nature of duality, the AdS/CFT correspondence provides a promising approach to studying gauge theories in strongly coupled regimes, where conventional perturbative techniques fail.
Beyond its wide applications in condensed matter physics~\cite{Hartnoll:2009sz,Herzog:2009xv,Cai:2015cya}, this paradigm has been extensively adopted in nuclear physics to investigate hadronic systems and hot/dense QCD matter~\cite{Erdmenger:2007cm,Casalderrey-Solana:2011dxg,Brodsky:2014yha,Brambilla:2014jmp}.
In particular, the five-dimensional Einstein-Maxwell-Dilaton (EMD) model, first proposed in~\cite{DeWolfe:2010he,DeWolfe:2011ts}, has opened new avenues for holographic studies of strongly coupled QGP.
Today, holographic EMD models are widely applied in diverse areas such as heavy quark potentials~\cite{Zhou:2020ssi}, transport properties~\cite{Rougemont:2017tlu,Grefa:2022sav}, and jet quenching~\cite{Rougemont:2015wca,Li:2025ugv}.
Recently, in the study of the QCD phase diagram~\cite{DeWolfe:2010he,DeWolfe:2011ts,He:2013qq,Yang:2014bqa}, improved holographic EMD models~\cite{Knaute:2017opk,Critelli:2017oub,Grefa:2021qvt,Cai:2022omk,Li:2023mpv,Rougemont:2023gfz}, calibrated against lattice QCD data at zero baryon chemical potential, have attracted significant interest for their ability to investigate the CEP and the first-order deconfinement transition.
In the bulk, the EMD model arises from a nonminimal coupling between a nonconformal dilatonic scalar field and a $U(1)$ gauge field.
Nonperturbative effects and flavor dynamics are effectively encoded into the model parameters by matching lattice QCD data, with a key challenge being the construction of the scalar potential $V(\psi)$ and the coupling function $f(\psi)$ along with the determination of their parameters. 
Holographically, the scalar field condensation drives conformal symmetry breaking, while the Abelian gauge field dual to the boundary theory's baryon number current introduces a finite chemical potential by generating an appropriate electric field in the black hole spacetime.

As mentioned earlier, the application of holographic EMD models to QCD physics via gauge/gravity duality has been extensively discussed, yet the investigation of black hole properties and spacetime geometry in the bulk remains relatively unexplored.
In the holographic superconductivity models~\cite{Gubser:2008px,Hartnoll:2008kx,Hartnoll:2008vx}, lowering the temperature to a critical point triggers the condensation of a scalar field that serves as an order parameter to simulate the superconducting phase transition.
On the gravity side, this mechanism manifests as instabilities of AdS black holes, which undergo transitions into hairy black holes.
Studying these black hole state transitions is crucial, as it provides direct insight into how phase structures in the dual field theory map onto phenomena in weakly coupled systems.
This understanding can, in turn, guide refinements of holographic models aimed at more faithfully capturing the QCD phase structure.
In~\cite{Guo:2024ymo}, we demonstrated that the holographic EMD model exhibits remarkable structural similarity to Einstein-Maxwell-scalar (EMS) models in the context of spontaneous scalarization.
Consequently, a simplified EMD model without lattice QCD data matching has been employed to investigate bulk instabilities and the associated hairy black hole solutions.
Our results revealed two distinct hairy black holes in the bulk, arising respectively from the scalar potential $V(\psi)$ and from the gauge field coupling function $f(\psi)$.
Through analyzing these black hole states, we identified two corresponding phases and their phase boundary in the dual field theory.
Thermodynamic investigations along this phase boundary not only recovered phase structures qualitatively consistent with holographic EMD models that a first-order phase transition terminating at a critical point, but also uncovered a third-order phase transition beyond this critical point.

Motivated by insights from the study of hairy black holes in the simplified EMD model within black hole physics, we extend our investigation in this work to the improved holographic EMD models proposed in~\cite{Critelli:2017oub,Grefa:2021qvt}, with a particular focus on hairy black hole solutions and their dual phase structures.
The model incorporates a real scalar field with potential $V(\psi)$ nonminimally coupled to the Maxwell field through a coupling function $f(\psi)$.
Two distinct hairy black hole solutions are discovered in this model, which through boundary condition discussions can be attributed to contributions from the potential function $V(\psi)$ and coupling function $f(\psi)$ respectively.
 Following an approach similar to~\cite{Guo:2024ymo}, our black hole physics analysis enables a mapping of the phase distributions and the corresponding phase boundary in the $(\mu_B,T)$ plane of the dual theory.
 Thermodynamic studies along this phase boundary reveal the coexistence of a first-order phase transition and a third-order phase transition lines terminating at a critical point.

The remainder of this paper is organized as follows.
In Sec.~\ref{sec=model}, we introduce the holographic EMD model under consideration and derive the corresponding equations of motion along with the necessary boundary conditions.
Sec.~\ref{sec=solution} presents the numerical results for static black hole solutions, where we construct the hairy black hole configurations and explore their domain of existence as well as their key physical properties.
In Sec.~\ref{sec=phase}, we study the phase diagram of these two hairy black hole solutions and discuss the phase transition along the phase boundary thorough Gibbs conditions.
Finally, Sec.~\ref{sec=conclusion} is devoted to further discussions and concluding remarks.

\section{Einstein-Maxwell-dilaton Model}\label{sec=model}

Now, this paper begin with a five-dimensional EMD model in asymptotically AdS spacetime
\begin{align}\label{eq=action}
	S=\frac{1}{2\kappa_5^2}\int d x^5\left(R-\frac{1}{2} \nabla^\mu \psi \nabla_\mu \psi-\frac{1}{4} f(\psi) F_{\mu \nu}^2 - V(\psi)\right),
\end{align}
where $\kappa_5^2\equiv8\pi G_5$ is the five-dimensional Newton's constant, $R$ represents the Ricci scalar, $\nabla_\mu$ denotes the covariant derivative, and $F_{\mu\nu}\equiv\nabla_\mu A_nu-\nabla_\nu A_\mu$ is the Maxwell tensor. 
A real massive scalar field $\psi$ is nonminimally coupled to the Maxwell field via the coupling function
\begin{align}
	f(\psi)=\frac{1}{1+c_3}\left(\sech(c_1\psi+c_2\psi^2)+c_3 \sech(c_4\psi) \right),
\end{align}
while the potential function reads
\begin{align}
	V(\psi)=-12\cosh(v_1\psi)+v_2\psi^2+v_3\psi^4+v_4\psi^6.
\end{align}
The free parameters in the functions $f(\psi)$ and $V(\psi)$ can be determined through calibration against state-of-the-art lattice QCD results for 2+1 flavor systems with physical quark masses at $\mu_B=0$. 
By holographically matching both the equation of state and the second-order baryon susceptibility $\chi_2^B$, the specific values of these parameters can be obtained.
In this work, we adopt the results from~\cite{Grefa:2021qvt}, as $c_1=-0.27,c_2=0.4,c_3=1.7,c_4=100, v_1=0.63,v_2=0.65,v_3=-0.05,v_4=0.003$ and $\kappa_5^2=8\pi G_5=8\pi(0.46)$.

The variation of the action~\eqref{eq=action} with respect to the matter and gravitational fields yield the following equations of motion
\begin{align}
   & \nabla_\mu \nabla^\mu \psi-\frac{\partial_\psi f}{4} F_{\mu \nu} F^{\mu \nu}-\partial_\psi V =0, \label{eq=scalar}\\
   & \nabla^\mu\left(f(\psi)F_{\mu\nu}\right)=0,\label{eq=maxwell}\\
   & R_{\mu \nu}-\frac{1}{2} R g_{\mu \nu}=\frac{1}{2}\left(\nabla_\mu \psi \nabla_\nu \psi+f(\psi) F_{\mu \rho} F_\nu{ }^\rho\right)-\frac{1}{2}g_{\mu\nu}\left(\frac{1}{2} \nabla_\rho\psi \nabla^\rho \psi+\frac{1}{4} f(\psi) F_{\sigma \rho}^2+V(\psi)\right),\label{eq=metric}
\end{align}
where we define $\partial_\psi f\equiv\frac{\partial f(\psi)}{\partial\psi}$ and $\partial_\psi V\equiv\frac{\partial V(\psi)}{\partial\psi}$.
In the following study, we consider a five-dimensional static translationally invariant black hole background, described by a general ansatz
\begin{align}
	ds^2 & = e^{2A(r)}\bigg[-g(r)dt^2 +d\vec{x}^{\ 2}\bigg]+\frac{e^{2B(r)}}{g(r)}dr^2,\\
	A & = \phi(r)dt,\\
	\psi & = \psi(r).
\end{align}
The event horizon is given by $g(r_h)=0$ and the asymptotically $\text{AdS}_5$ boundary is located at $r\rightarrow\infty$.
As a result, the field equations can be written as 
\begin{align}
	& \psi''(r) + \left(\frac{g'(r)}{g(r)}+4A'(r)-B'(r)\right)\psi'(r) - \frac{e^{2B(r)}}{g(r)}\left(\partial_\psi V-e^{-2(A(r)+B(r))}\frac{\partial_\psi f}{2}\phi'(r)^2\right)=0, \label{eq=motion1}\\
	& \phi''(r)+ \left(2A'(r)-B'(r)+\frac{\partial_\psi f}{f(\psi)}\psi'(r)\right)\phi'(r)=0, \label{eq=motion2}\\
	& g''(r) + \left(4A'(r)-B'(r)\right) - e^{-2A(r)}f(\psi) \phi'(r)^2 =0, \label{eq=motion3}\\
	& 2 A''(r) - A'(r)B'(r) + \frac{\psi'(r)^2}{6} = 0, \label{eq=motion4}\\
	& g(r)\left(24A'(r)^2-\psi'(r)^2\right) + 6A'(r)g'(r) + 2e^{2B(r)}V(\psi) + e^{-2A(r)}f(\psi) \phi'(r)^2 = 0. \label{eq=motion5}
\end{align}
Notably, the last equation~\eqref{eq=motion5} is an independent constraint for the metric fields.
From the above equations of motion, it is convenient to derive the Gauss charge $Q_G$ and Noether charge $Q_N$ with respect to the first integral of the Maxwell field~\eqref{eq=motion2} and metric field~\eqref{eq=motion3}, respectively
\begin{align}
	& Q_G=f(\psi)e^{2A(r)-B(r)}\phi'(r),\label{eq=QG}\\
	& Q_N=e^{2A(r)-B(r)}\left(e^{2A(r)}g'(r)-f(\psi)A_t'(r)\phi(r)\right).\label{eq=QN}
\end{align}

Since the background function $B(r)$ does not contain any dynamical information, we can impose the gauge condition $B(r) = 0$ to simplify numerical computations.
Consequently, in our model framework, the obtained field functions include $\psi(r), \phi(r), g(r)$, and $A(r)$, along with their corresponding system of ordinary differential equations.
Moreover, the absence of explicit radial coordinate $r$ dependence in both the ansatz and field equations implies that we can always rescale the horizon position to $r_h=0$ through the coordinate transformation $r\rightarrow r-r_h$.
Therefore, to numerically solve the above equations, we impose the following boundary conditions for each field function near the horizon
\begin{align}
	\psi(r\rightarrow 0) &= \psi_h^0+\psi_h^1 r + \psi_h^2 r^2 +\dots,\label{eq=bdhorizon1}\\ 
	\phi(r\rightarrow 0) &= \phi_h^1 r+ \phi_h^2 r^2 +\dots,\label{eq=bdhorizon2}\\ 
	g(r\rightarrow 0)    &= g_h^1 r+ g_h^2 r^2 +\dots, \label{eq=bdhorizon3}\\ 
	A(r\rightarrow 0)    &= A_h^0 + A_h^1 r+ A_h^2 r^2 +\dots . \label{eq=bdhorizon4}
\end{align}
The solutions of black hole configuration require the blackening function vanish near the event horizon as $g(r\rightarrow 0)=0$.
$g_h^1=1$ can be fixed by considering the scaling symmetry of the time coordinate, and $A_h^0=0$ can be attained by rescaling the space and time coordinates simultaneously.
The Maxwell field also satisfies $\phi(r\rightarrow 0)=0$ because of the regular condition.
With these considerations, when substituting the above boundary conditions into the equations of motion, all remaining coefficients are determined by two independent parameters $\psi_h^0$ and $\phi_h^1$
\begin{align}
	& \psi_h^1=-\frac{1}{2}{\phi_h^1}^2f'(\psi_h^0)+V'(\psi_h^0),\label{eq=psi1}\\
	& \psi_h^2=\frac{1}{16}\left(-4\phi_h^1\big[V(\psi_h^0)f(\psi_h^0\big]'+\frac{{\phi_h^1}^2}{f(\psi_h^0)}\bigg[f(\psi_h^0)f''(\psi_h^0)-2f'(\psi_h^0)^2\bigg]\beta_{-}^{\prime}-2\beta_{-}^{\prime}V''(\psi_h^0)\right),\\
	& A_h^1=-\frac{1}{6}\beta_+,\quad A_h^2=-\frac{1}{48}{\beta_{-}^{\prime}}^2,\\
	& g_h^2=\frac{1}{6}\left(5{\phi_h^1}^2f(\psi_h^0)+4V(\psi_h^0)\right), \quad   \phi_h^2=\frac{1}{12}\phi_h^1\left(2\beta_{+}+\frac{3f'(\psi_h^0)}{f(\psi_h^0)}\beta_{-}^{\prime}\right),
\end{align}
with the coefficients $\beta_{\pm}\equiv {\phi_h^1}^2f(\psi_h^0)\pm 2V(\psi_h^0)$ and the prime symbol $^\prime=\frac{\partial}{\partial\psi_h^0}$.

The ultraviolet behaviors of these field functions at spatial infinity can be obtained to
\begin{align}
    A(r\rightarrow\infty) &\equiv \chi(r)=A_\infty^0+A_\infty^1 r+\mathcal{O}\left(e^{-2 \Delta\chi(r)}\right), \label{eq=inf_metric1}\\ 
    g(\rightarrow\infty) &=g_\infty^0+\mathcal{O}\left(e^{-4 \chi(r)}\right),   \label{eq=inf_metric2}\\
    \psi(\rightarrow\infty) &=\psi_\infty^0 e^{-\Delta\chi(r)}+\mathcal{O}\left(e^{-2 \Delta\chi(r)}\right), \label{eq=inf_scalar}\\
    \phi(\rightarrow\infty) &=\phi_\infty^0+\phi_\infty^2 e^{-2 \chi(r)}+\mathcal{O}\left(e^{-(2+\Delta) \chi(r)}\right),\label{eq=inf_matter}
\end{align}
According to the AdS/CFT correspondence, $\psi_\infty^0$ represents the asymptotic value of the scalar field, as the scaling dimension of the gauge theory operator is given by $\Delta_{\pm}=\frac{d\pm\sqrt{d^2+4m^2}}{2}$.
Due to exponential decay, the suppression of the term related to $\Delta_+$ is more significant than that of $\Delta_-$.
As a result, the asymptotic behavior of the scalar field remain $\Delta_-$ term as the source and we rewrite $\Delta=\Delta_-$.
Notably, considering the week approximation near the boundary, the potential function reduces to $V(\psi)=-12+\frac{1}{2}m^2\psi^2$ with the scalar mass defined by $m^2=-12v_1^2+2v_2$. 
This implies that the cosmological constant is given by $\Lambda=-6$ so the AdS radius is given to $L=1$.

The temperature and entropy density of the gauge theory fluid are holographically associated with the Hawking temperature and the area of the black hole horizon, respectively,
\begin{align}
	T=\frac{h'(0)}{4\pi},\quad s=\frac{2\pi}{\kappa_5^2}.
\end{align}
The baryon chemical potential and charge density are obtained from the boundary value of the gauge field
\begin{align}
	\mu_B=\lim_{r\rightarrow\infty}\phi(r)=\phi_\infty^0,\quad \rho_B=\lim_{r\rightarrow\infty}\frac{Q_G}{\kappa_5^2}=-\frac{\phi_\infty^2}{\kappa_5^2}.
\end{align}

Before proceeding with numerical computations, it is important to note that, as pointed out in~\cite{Grefa:2021qvt,Critelli:2017oub}, all preceding analyses are conducted in numerical coordinates for computational convenience. 
To extract physical quantities in the dual gauge theory within the holographic framework, one need to rescale the results to standard coordinates. 
Following the methodology in~\cite{Cai:2022omk}, the source $\psi_\infty^0$ of the scalar operator serves as the energy scale, allowing the rescaling of all physical quantities in the gauge theory so that they can match the QCD-scale holographic system. 
Although these two approaches differ in their operational implementations, they prove equivalent in their physical predictions. 
In~\cite{Grefa:2021qvt,Critelli:2017oub}, the adopted energy scale $\Lambda_\psi=1058.83 \text{MeV}$ leading to an actual energy rescaling factor
\begin{align}
	\lambda_\psi=\frac{\Lambda_\psi}{(\psi_\infty^0)^{1/\Delta}}.
\end{align}
Alternatively, if the asymptotic form of the blackening function $g(r)$ is rescaled to unity, this introduces the coordinate rescaling transformation $t\rightarrow \sqrt{g_\infty^0}t$ and $r\rightarrow\frac{r}{\sqrt{g_\infty^0}}$. 
Incorporating these perspectives, the physical quantities in the dual theory can be expressed as
\begin{align}\label{eq=thermoquan}
    & T =\frac{\lambda_\psi}{4\pi\sqrt{g_\infty^0}}, \quad
    s =\frac{2 \pi}{\kappa_5^2}\lambda_\psi^3, \quad
    \mu_B =\frac{\phi_\infty^0}{\sqrt{g_\infty^0}}\lambda_\psi, \quad
    \rho_B =-\frac{\phi_\infty^2}{\kappa_5^2\sqrt{g_\infty^0}}\lambda_\psi^3.
\end{align}

\section{Numerical Solutions}\label{sec=solution}

\subsection{Numerical procedure}

In this section, we numerically solve the aforementioned field equations. 
Guided by boundary conditions (\ref{eq=bdhorizon1}-\ref{eq=bdhorizon4}), the numerical integration initiates near the event horizon. 
To circumvent the singularities, we set the initial integration point at $r_{\text{start}}=10^{-8}$. 
The integration terminates at a cutoff radius $r_{\text{end}}=10$, where the field functions' exponential decay asymptotics ensure numerical results already satisfy boundary requirements~(\ref{eq=inf_metric1}-\ref{eq=inf_matter}) at spatial infinity. 
As previously established, all degrees of freedom in the dual theory are governed by two independent parameters $(\psi_h^0,\phi_h^1)$. 
Through constraint condition~\eqref{eq=motion5}, the gauge field parameter $\phi_h^1$ are found to obey the following bound
\begin{align}\label{eq=bound}
	\phi_h^1<\sqrt{-\frac{2V(\psi_h^0)}{f(\psi_h^0)}}\equiv\Phi_1^{\text{max}}(\psi_h^0),
\end{align}
in order to ensure $A_h^1>0$ in the model.

Upon completing the numerical integration to the cutoff boundary, we proceed to extract coefficients of all field functions in the far-field asymptotic region. 
The parameters $g_\infty^0=g(r_{\text{end}})$ and $\phi_\infty^0=\phi(r_{\text{end}})$ can be directly determined from the field profiles at radial cutoff. 
Implementing constraint condition~\eqref{eq=motion5}, which establishes $A_\infty^1=1/\sqrt{g_\infty^0}$, allows subsequent derivation of $A_\infty^0=\chi(r_{\text{end}})-A_\infty^1\times r_{\text{end}}$ through the asymptotic behavior of metric function.
Furthermore, by enforcing the conserved charge~\eqref{eq=QG} evaluated at both the horizon and the boundary, the value of $\phi_\infty^2$ is constrained as
\begin{align}
	\phi_\infty^2=-\frac{\sqrt{g_\infty^0}}{2f(0)}f(\psi_h^0)\phi_h^1.
\end{align}
The remaining parameter, $\psi_\infty^0$, is subsequently determined by matching the scalar field's asymptotic profile in the far-field region~\cite{Critelli:2017oub}.

\begin{figure}[thbp]
    \centering
    \includegraphics[width=0.48\linewidth]{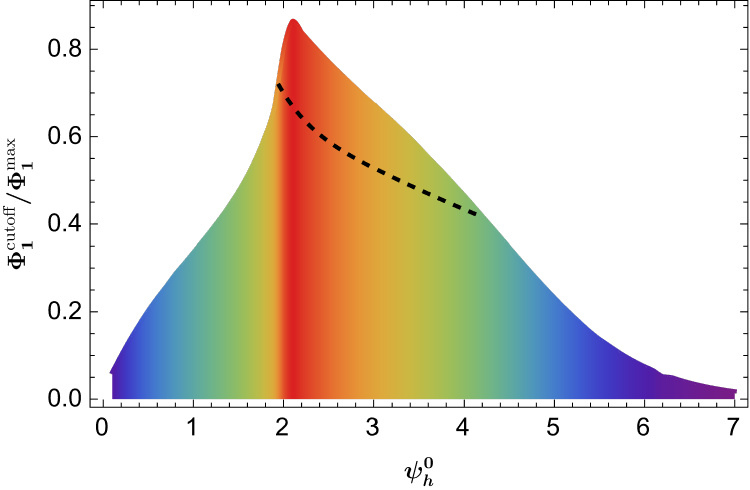}
    \includegraphics[width=0.48\linewidth]{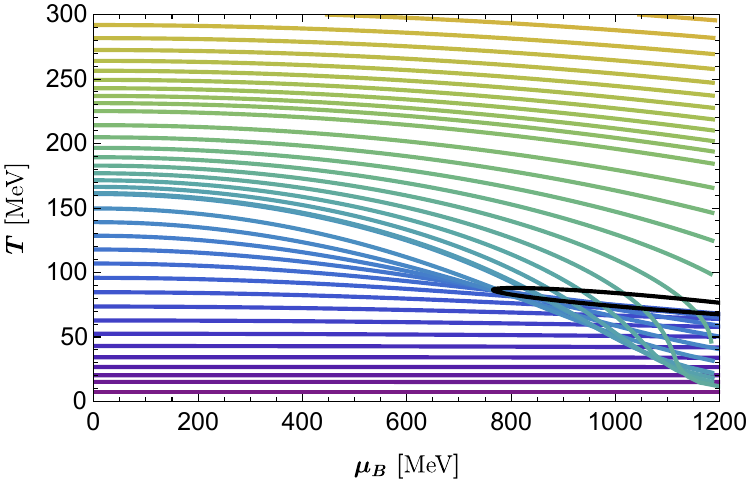}
    \caption{Left: Existence domain of the hairy black holes in the parameter space of $(\psi_h^0,\Phi_1^{\text{cutoff}}/\Phi_1^{\text{max}})$, the black dashed line represents the phase boundary between two different types of hairy solutions; Right: Hairy black hole solutions for different $\psi_h^0$ in the parameter space of $(\mu_B,T)$, the black solid curve describes the phase boundary as mentioned in the left panel.}
    \label{fig=parameter}
\end{figure}

With the asymptotic boundary coefficients of all field functions now obtained through the procedures outlined above, the relevant physical quantities~\eqref{eq=thermoquan} in the dual theory can be determined accordingly.
To this end, it is necessary to compute hundreds of thousands of black hole solutions to cover most of the QCD phase diagram, particularly covering the primary region of interest with $\mu_B<1200\text{MeV}$ and $T<800\text{MeV}$.
However, the bound specified in Eq.~\eqref{eq=bound} would computationally over-sample parameter space beyond physically relevant domains.
Exploiting the monotonic relationship between $\mu_B$ and $\phi_h^1$ for fixed $\psi_h^0$, we implement an optimized protocol by introducing a cutoff value $\Phi_1^{\text{cutoff}}$ around $\mu_B\approx 1200\text{MeV}$, as illustrated in the left panel of Fig.~\ref{fig=parameter} where color-coding denotes our sampled parameter space.
In the right panel, we present representative data showing the relationship between the temperature $T$ and the chemical potential $\mu_B$ for various values of $\psi_h^0$.
Each curve corresponds to a fixed $\psi_h^0$, with increasiing $\phi_h^1$ tracking the growth of $\mu_B$.
It is evident that, in the low-temperature and high-chemical potential region, the curves overlap, indicating the existence of competing black hole solutions—precisely the region where the holographic QCD phase transition occurs.
From the bulk perspective, this degeneracy reflects the competition among different types of black hole states, motivating a detailed analysis of these solutions across various regions of the phase space.

\subsection{Black hole solutions in the bulk}

\begin{figure}[thbp]
    \centering
    \includegraphics[width=0.48\linewidth]{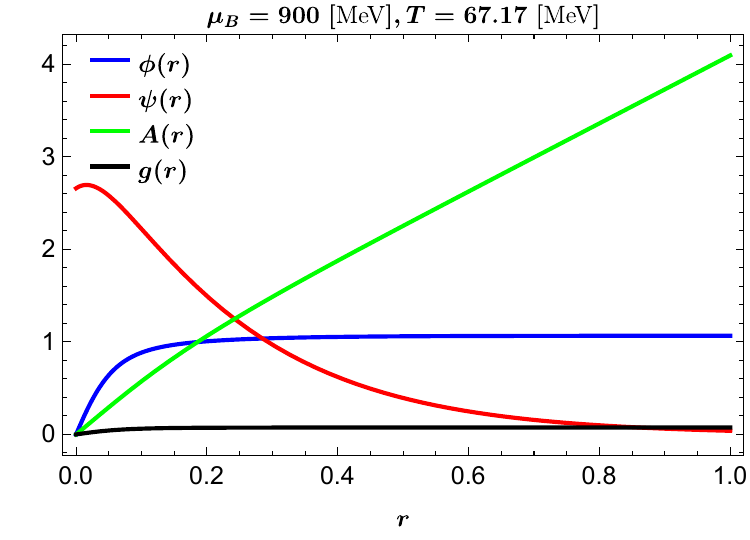}
    \includegraphics[width=0.48\linewidth]{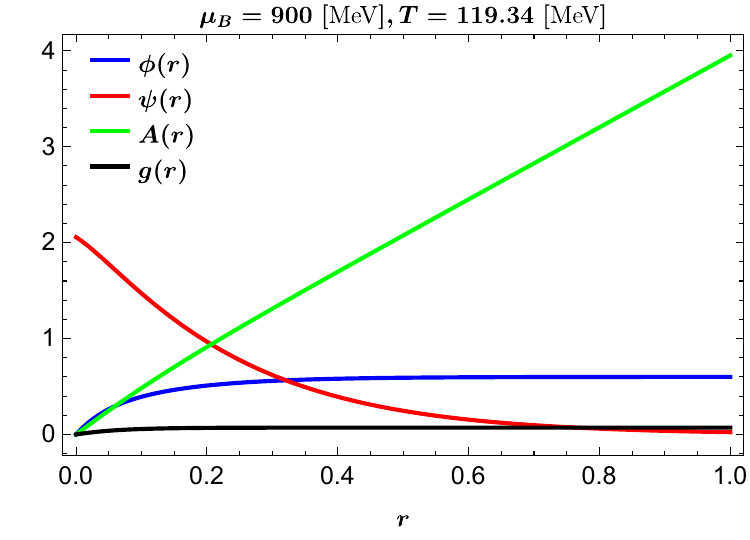}
    \caption{A typical example of the profiles of the field functions corresponding to Type II hairy black hole solutions (left) and Type I hairy black hole solutions (right). }
    \label{fig=profile}
\end{figure}

We now examine the radial profiles of the various field functions in the bulk as obtained from the numerical solutions.
Based on our previous discussion of the right panel of Fig.~\ref{fig=parameter}, which suggested potential competition between two distinct black hole solutions, the profile analysis further confirms that their primary difference manifest in the scalar field's spatial evolution.
As a representative example, we examine the black hole solutions and the associated field profiles at a baryon chemical potential $\mu_B=900\text{MeV}$ for different temperatures.
The left panel of Fig.~\ref{fig=profile} corresponds to a parameter point within the overlapping region depicted in the right panel of Fig.~\ref{fig=parameter}, whereas the right panel represents a point above this overlapping region in the parameter space.
In the non-overlapping region, the scalar field $\psi(r)$ exhibits monotonic decay along the radial coordinate, a behavior observed in various models of hairy black hole solutions~\cite{Fernandes:2019rez,Herdeiro:2018wub,Hertog:2004dr}.
In contrast, within the overlapping regime, the scalar field displays nonmonotonic behavior, as it shows an initial growth near the horizon followed by smooth asymptotic decay after reaching a local maximum.

Numerically, the near-horizon behavior of the scalar field profile is determined by the boundary conditions imposed on the equations of motion.
In particular, the growth/decay tendency at $r\approx r_h$ relies critically on the sign of $\psi_h^1$ in Eq.~\eqref{eq=psi1}.
A negative $\psi_h^1$ produce exclusively decaying profiles, whereas a positive $\psi_h^1$ induce the scalar field amplification, effectively forming a barrier near the horizon.
Eq.~\eqref{eq=psi1} explicitly reveals the competition between the coupling function $f(\psi)$ and the scalar potential $V(\psi)$ through their first derivatives.
Since these derivatives enter the constraint with a negative sign, dominance of $V'(\psi_h^0)$ leads to a negative $\psi_h^1$, driving monotonic decay as illustrated in the right panel of Fig.~\ref{fig=profile}.
Conversely, when $f'(\psi_h^0)$ dominates, $\psi_h^1$ becomes positive, yielding the non-monotonic behavior observed in the left panel of Fig.~\ref{fig=profile}.
These two distinct types of scalar field profiles are consistent with the findings of~\cite{Guo:2024ymo}: in models containing only a scalar potential, a massive scalar field typically exhibits a monotonically decaying profile, whereas a massless scalar field with nonminimal electromagnetic coupling develops monotonic growth and asymptotes to a constant at spatial infinity.

This system therefore hosts two competing hairy black hole states. 
For clarity, we classify solutions with monotonic decay as Type I hairy black holes, and those exhibiting nonmonotonic behavior as Type II.
The phase boundary between these two types is determined by the vanishing condition of Eq.~\eqref{eq=psi1}, indicated by the dashed line in the left panel of Fig.~\ref{fig=parameter}, where solutions above the curve correspond to the parameter space of the Type II hairy black holes. 
Traversing this boundary by tuning $\psi_h^0$ and $\phi_h^1$ induces a transition between the two types of hairy black holes, which holographically corresponds to a phase transition through the $T-\mu_B$ variations in the dual field theory.

\begin{figure}[thbp]
    \centering
    \includegraphics[width=0.6\linewidth]{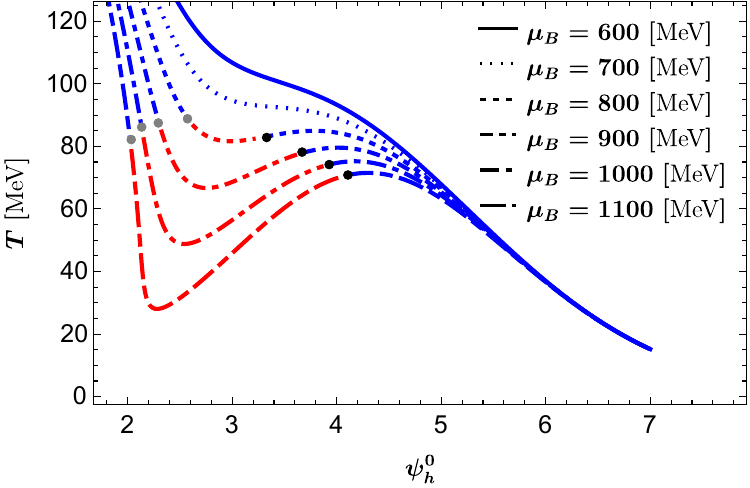}
    \caption{Temperature as a function of the scalar hair $\psi_h^0$ with different $\mu_B$. 
    The blue curves represent the type I hairy states, while the red segments correspond to the type II hairy black holes. 
    The left intersection points between these two hairy states are denoted by the gray dots which correspond to the upper bifurcation of phase boundary, while the right intersection points are marked by the black dots which are associated with the lower bifurcation of the phase boundary in the right panel of Fig~\ref{fig=parameter} and Fig.~\ref{fig=phase}.}
    \label{fig=Tvpsi0}
\end{figure}

The phase boundary between the two black hole states in the $T-\mu_B$ parameter space is represented by the solid black curve in the right panel of Fig.~\ref{fig=parameter}. 
Unlike the dashed line in the left panel, this boundary forms a U-shaped structure in the phase space.
The region outside the boundary corresponds to the Type I hairy solutions; however, this does not imply that the Type II hairy black hole phase exists only within the boundary, as it also encompasses the overlapping region below the boundary curve.
Notably, the lower segment of this phase boundary coincides with the edge of the overlapping region, where a first-order QCD phase transition occurs~\cite{Critelli:2017oub,Grefa:2021qvt}.
Our findings further identify an upper boundary segment, suggesting the presence of additional phase transitions, the nature of which will be explored in subsequent sections.

\begin{figure}[thbp]
    \centering
    \includegraphics[width=0.48\linewidth]{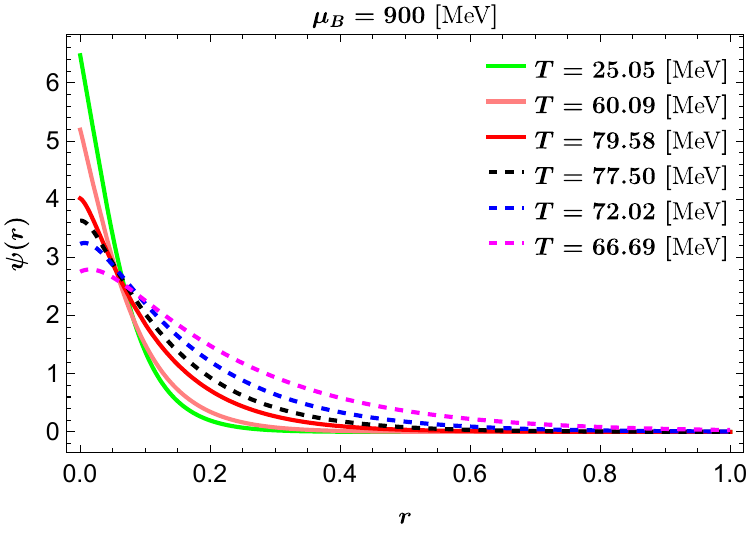}
    \includegraphics[width=0.48\linewidth]{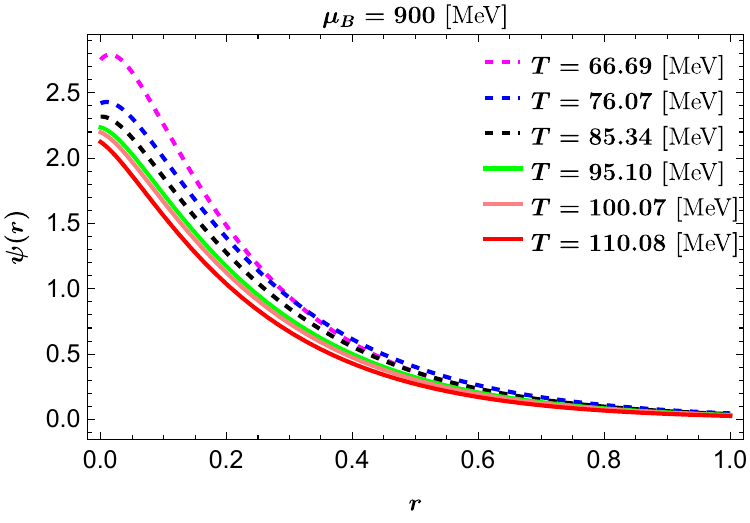}
    \caption{The radial profiles for scalar field with different temperature at $\mu_B=900\text{MeV}$. 
    The solid profiles represents the type I hairy solutions, the dashed curves denote type II hairy solutions. 
    The black dashed line indicates the solutions at the intersection point.}
    \label{fig=scalar}
\end{figure}

The phase boundary reveals that the Type II hairy black hole state exhibits nontrivial behavior in the phase space.
To elucidate this complex phase behavior, we fix different values of the chemical potential and examine the relationship between the temperature $T$ and the scalar hair $\psi_h^0$.
The results are shown in Fig.~\ref{fig=Tvpsi0}, where the blue curves represent the Type I hairy black hole solutions, and the red segment corresponds to the Type II hairy state.
Below a critical value of $\mu_B$, only the Type I state exists, with temperature decreasing monotonically as the scalar hair is enhanced.
Beyond this threshold, Type II solutions emerge, creating two intersection points with the Type I state that correspond to the bifurcations of the U-shaped boundary in the right panel of Fig.~\ref{fig=parameter} at a fixed $\mu_B$.
Clearly, the Type II state exhibits nontrivial behavior that leads to an S-shaped curve for the entire $T-\psi_h^0$ functions.
Consistent with~\cite{Critelli:2017oub, Grefa:2021qvt}, the right intersection points (black dots) in Fig.~\ref{fig=Tvpsi0} correspond to first-order phase transitions in the holographic QCD phase diagram, i.e. the boundary of the overlapping region in the right panel of Fig.~\ref{fig=parameter}.
A detailed thermodynamic analysis is required to elucidate the physical significance of the left intersection points (gray dots), which are associated with the upper bifurcation of the phase boundary.

The scalar field profiles provide microscopic insights into the transitions between the two different types of hairy black hole states.
Fig.~\ref{fig=scalar} depicts solutions at $\mu_B = 900 \text{MeV}$, where solid (dashed) curves represent Type I (II) states, and black dashed lines denote the solutions at the intersection point.
The left panel illustrates the behavior of the scalar field at various temperatures near the right intersection point of Fig.~\ref{fig=Tvpsi0}. 
Type I solutions show diminishing scalar hair with increasing $T$, abruptly transitioning to Type II beyond the critical point with a reversed $T$-dependence.
This sharp crossover suggests a strong phase transition in the dual theory~\cite{Critelli:2017oub,Grefa:2021qvt}.
Conversely, the right panel shows that the scalar hair decreases gradually while the temperature increases continuously, indicating a smooth transition between the two black hole states.
This behavior suggests that, if a phase transition holographically occurs in this region, it would be a relatively mild process.

\section{Thermodynamics at finite chemical potential}\label{sec=phase}

Building upon the preceding analysis of bulk black hole solutions and field profiles, two distinct classes of hairy black hole states can be clearly identified.
Type I solutions arise from dominant potential contributions $V(\psi)$, generating scalar field profiles with monotonic radial decay.
Conversely, Type II solutions emerge when coupling function $f(\psi)$ governs the dynamics, producing near-horizon scalar field growth that forms a barrier.
Through critical condition analysis of Eq.~\eqref{eq=psi1}, we have mapped the phase boundary separating these two states, which clearly delineates their distribution in the parameter space.
Consequently, we now focus on analyzing the thermodynamic phase transition between these two distinct hairy phases in the dual boundary field theory.

\subsection{phase diagram}

\begin{figure}[thbp]
    \centering
    \includegraphics[width=0.6\linewidth]{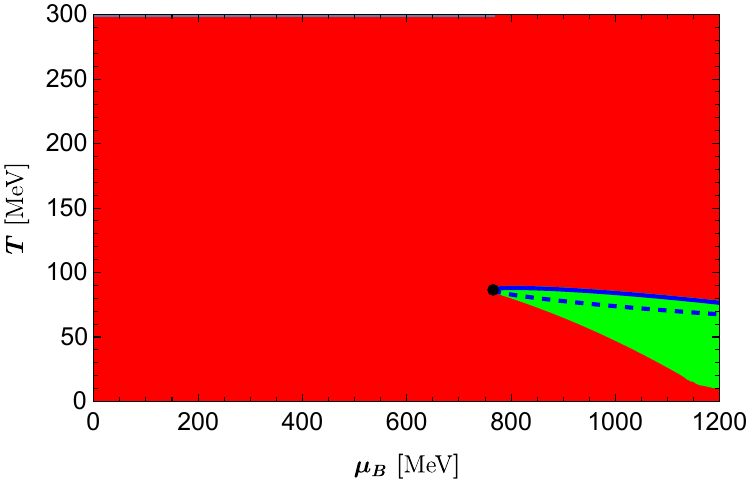}
    \caption{The phase diagram of these two hairy phases in parameter space of $(\mu_B,T)$. 
    The U-shaped blue curve represents the phase boundary between these two phases. 
    The type I hairy phase, denoted by the red region persist throughout the exterior domain of the boundary line, while the type II hairy phase is described by the green region. 
    The blue dashed segment of the phase boundary represents a first-order phase transition line, while the blue solid segment denotes a third-order phase transition line. 
    These two transition line terminates at a critical point $(\mu_B^{\text{crit}},T_{\text{crit}})=(765.51,86.54)\text{MeV}$ which is indicated by the black dot.}
    \label{fig=phase}
\end{figure}

In the dual field theory, a refined thermodynamic phase diagram in the $T-\mu_B$ parameter space is depicted in Fig.~\ref{fig=phase}.
Here, red and green regions denote Type I and Type II phase dominance respectively, bounded by the blue curve marking the critical phase boundary.
Notably, Type I solutions persist throughout the exterior domain of the boundary curve, while the green area below the blue dashed line represents the overlapping region where both phases coexist.

Comparisons with the results of holographic critical endpoint model~\cite{Critelli:2017oub,Grefa:2021qvt} reveal the blue dashed line, serving as the boundary of the overlapping region, coincides precisely with the line along which the first-order phase transition occurs in the holographic QCD theory.
As demonstrated in Fig.~\ref{fig=Tvpsi0}, this boundary exhibits sharp thermodynamic changes characteristic of strong phase transitions.
The solid blue segment represents the phase boundary between the two phases  located on the left side of Fig.~\ref{fig=Tvpsi0}.
In the next subsection, it will be demonstrated via the Gibbs conditions that a subtler third-order phase transition occurs between the two phases, whose weak signatures elude conventional holographic detection.
These two transition branches converge at a critical point $(\mu_B^{\text{crit}},T_{\text{crit}})=(765.51,86.54)\text{MeV}$, which exactly corresponds to the turning point of the boundary curve, marked in the figure by a solid black dot.
This phenomenon was first observed in a simplified EMD model that did not match lattice QCD data~\cite{Guo:2024ymo}, wherein the overlapping region between the two phases was much smaller and the critical point for the two phase transitions did not appear at the turning point of the phase boundary.
This indicates that by appropriately tuning the parameters and forms of the scalar potential $V(\psi)$ and the coupling function $f(\psi)$, one can modify the structure of the phase diagram, thereby opening up avenues for investigating more complex types of phase transitions from a gravitational perspective.

\subsection{Phase transition}

\begin{figure}[thbp]
    \centering
    \includegraphics[width=0.24\linewidth]{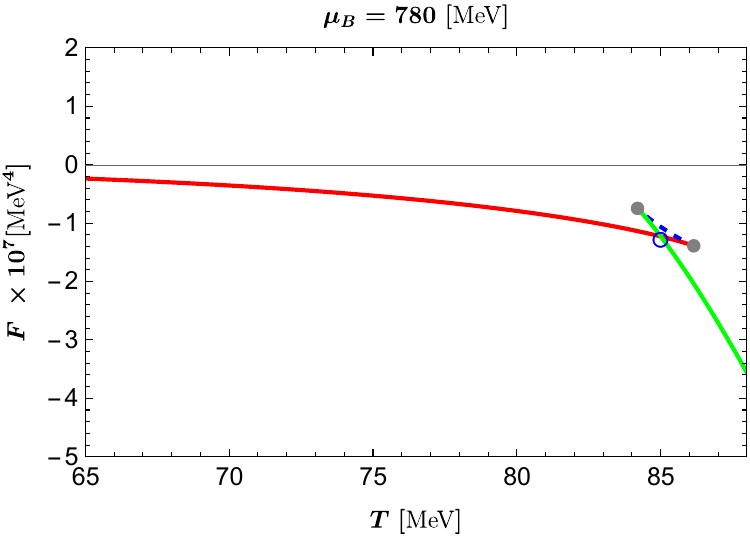}
    \includegraphics[width=0.24\linewidth]{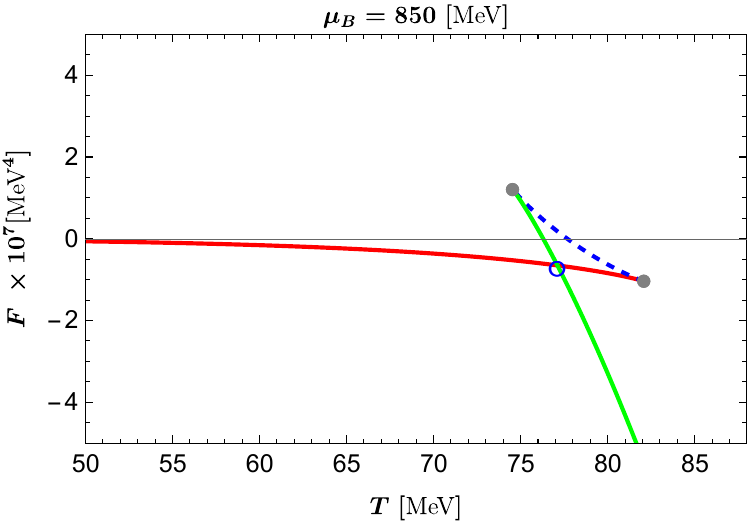}
    \includegraphics[width=0.24\linewidth]{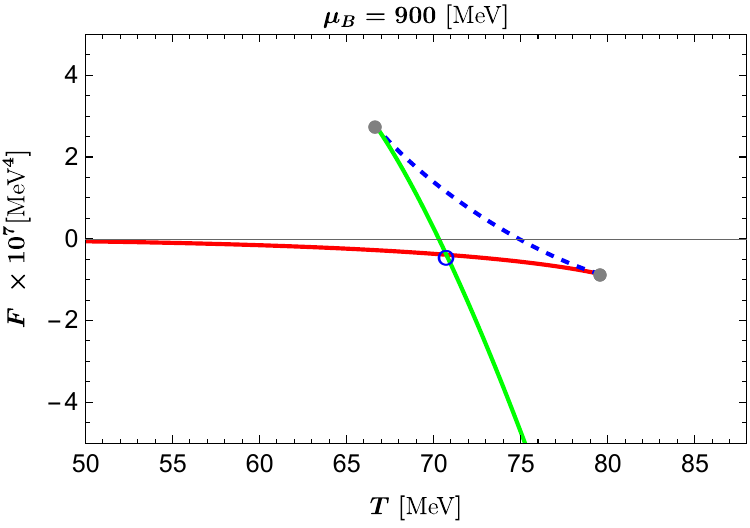}
    \includegraphics[width=0.24\linewidth]{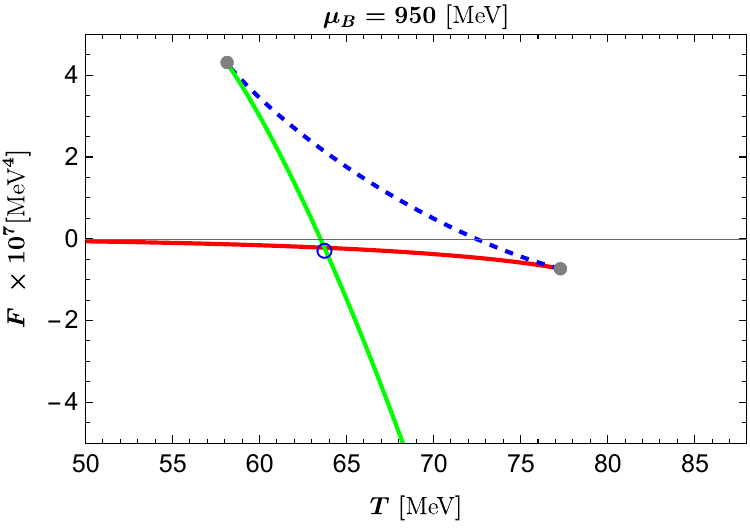}
    \includegraphics[width=0.24\linewidth]{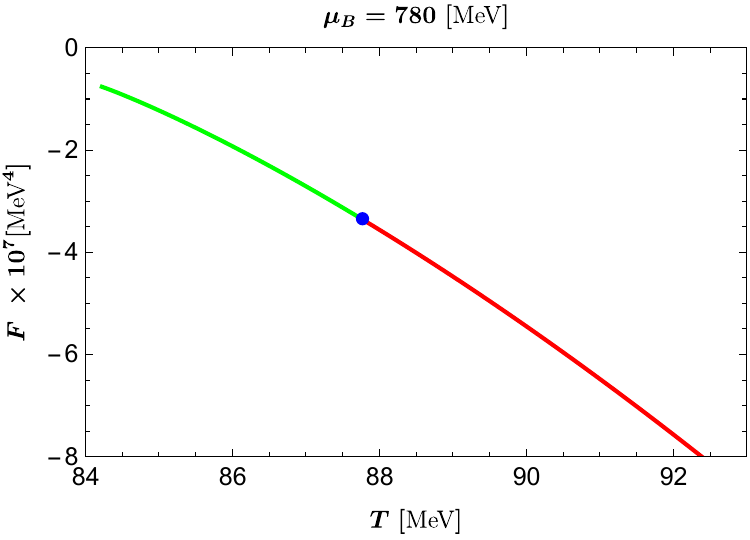}
    \includegraphics[width=0.24\linewidth]{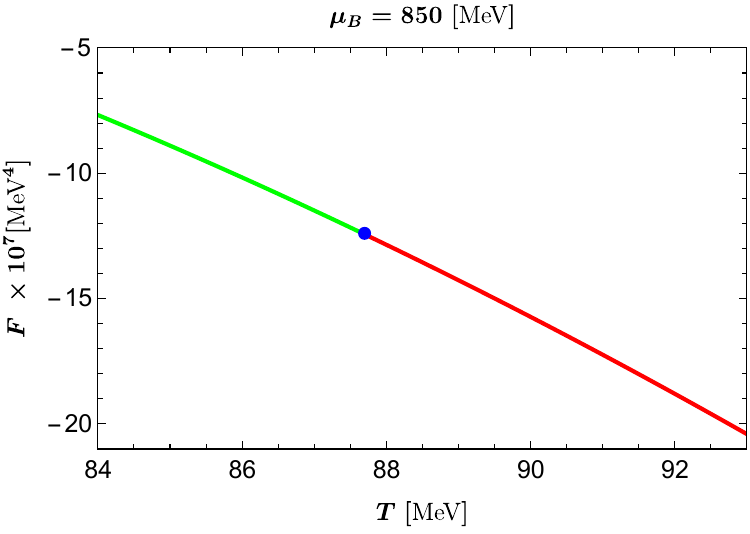}
    \includegraphics[width=0.24\linewidth]{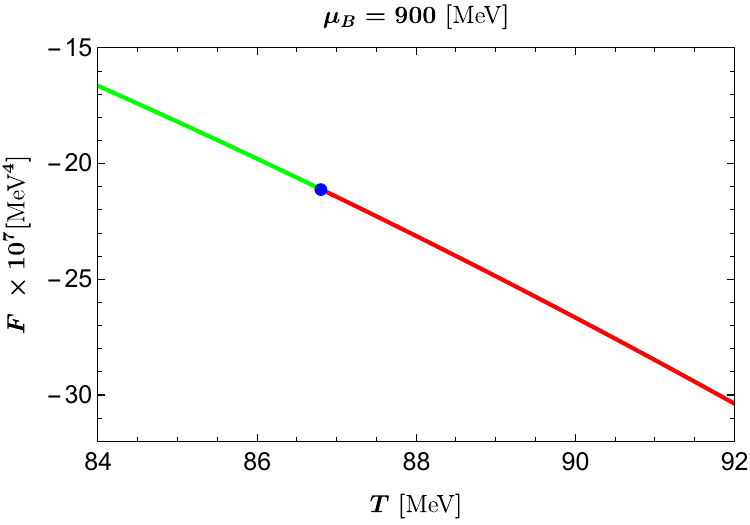}
    \includegraphics[width=0.24\linewidth]{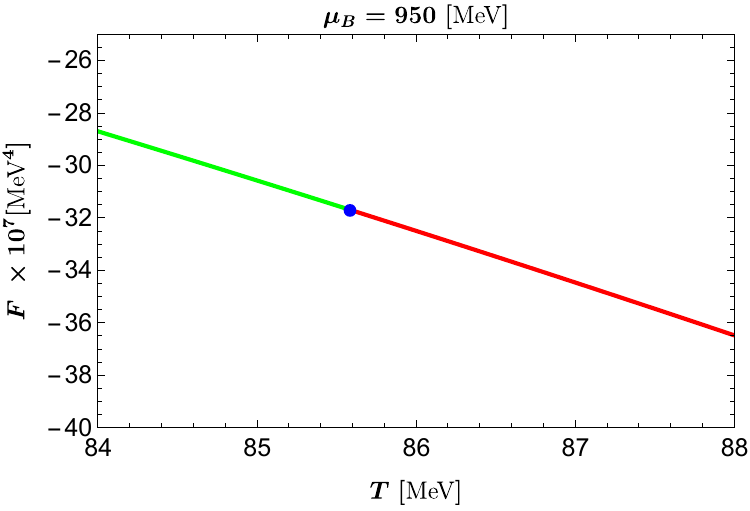}
    \caption{The Gibbs free energy density with respect to temperature for different chemical potentials.
    The red (green) solid lines represent the type I(II) hairy phases, and dashed blue curve corresponds to hairy black hole solutions that are thermodynamically unstable. 
    The top panels show phase trajectories that cross the first-order transition line, with the blue circle marking the transition point, whereas the bottom panels illustrate trajectories across the third-order transition line, where the transition point is indicated by a blue dot. 
    }
    \label{fig=freeenergy}
\end{figure}

\begin{figure}[thbp]
    \centering
    \includegraphics[width=0.24\linewidth]{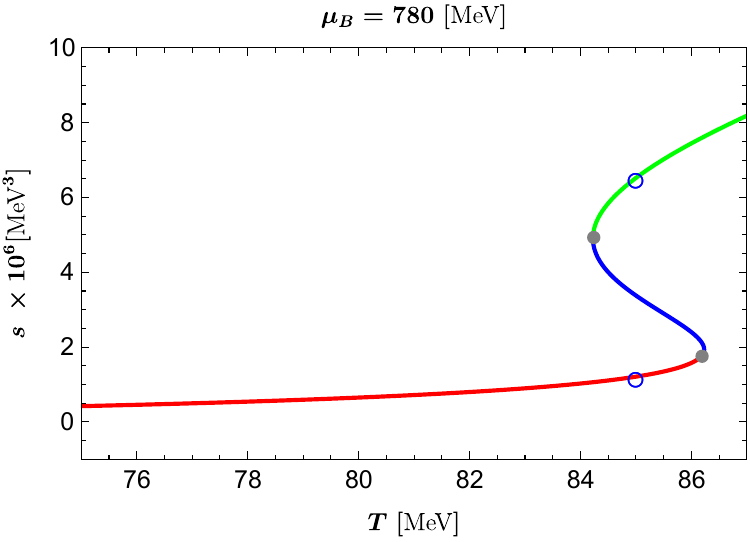}
    \includegraphics[width=0.24\linewidth]{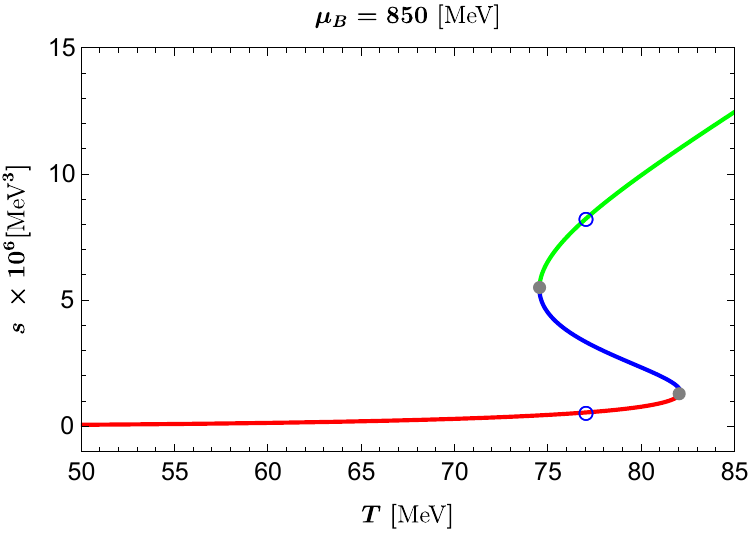}
    \includegraphics[width=0.24\linewidth]{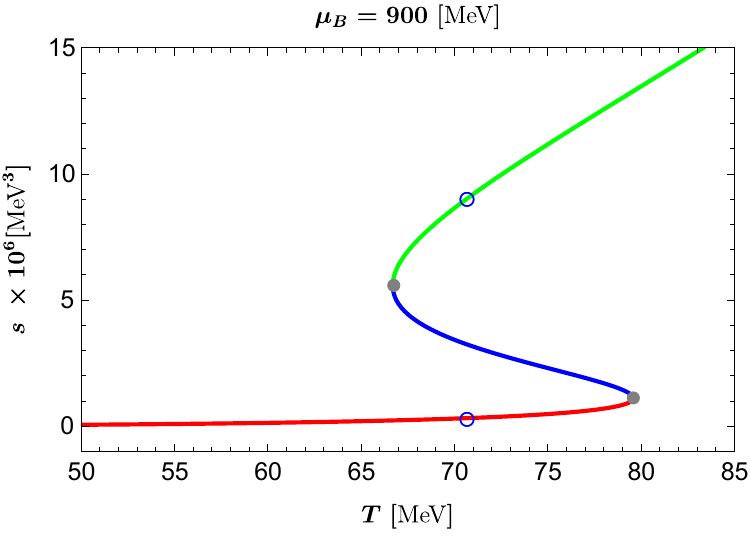}
    \includegraphics[width=0.24\linewidth]{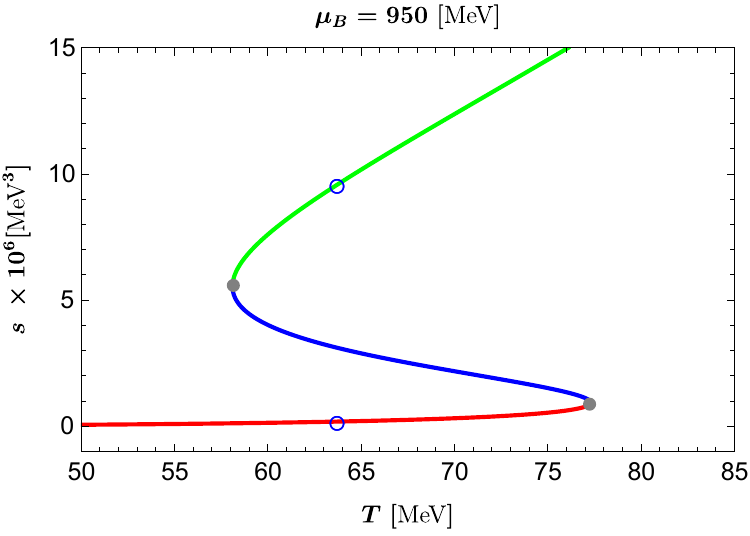}
    \includegraphics[width=0.24\linewidth]{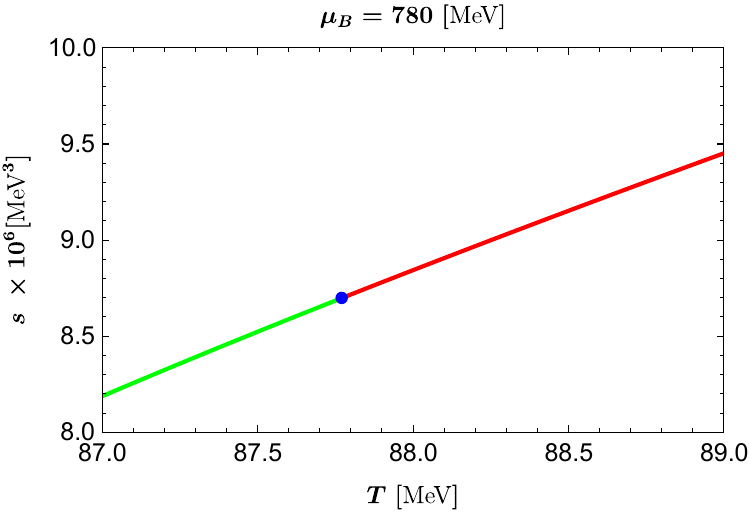}
    \includegraphics[width=0.24\linewidth]{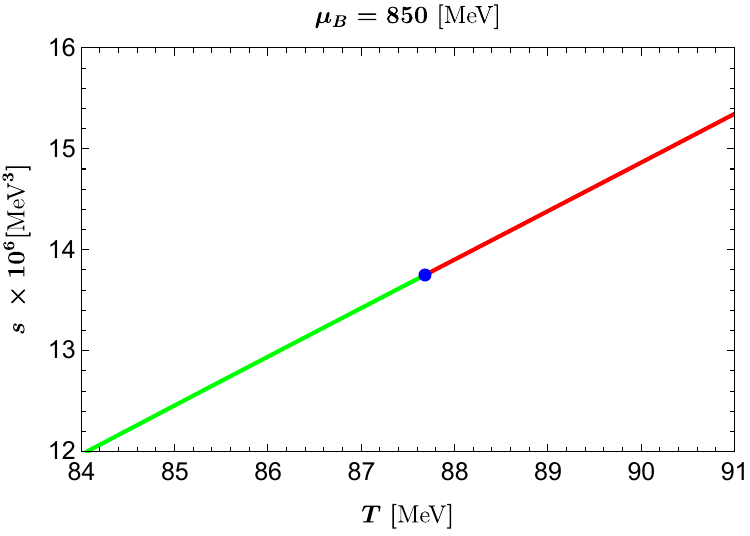}
    \includegraphics[width=0.24\linewidth]{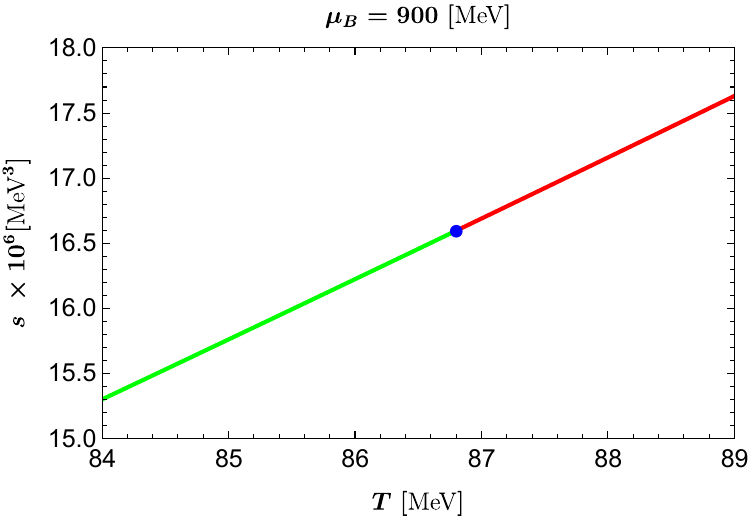}
    \includegraphics[width=0.24\linewidth]{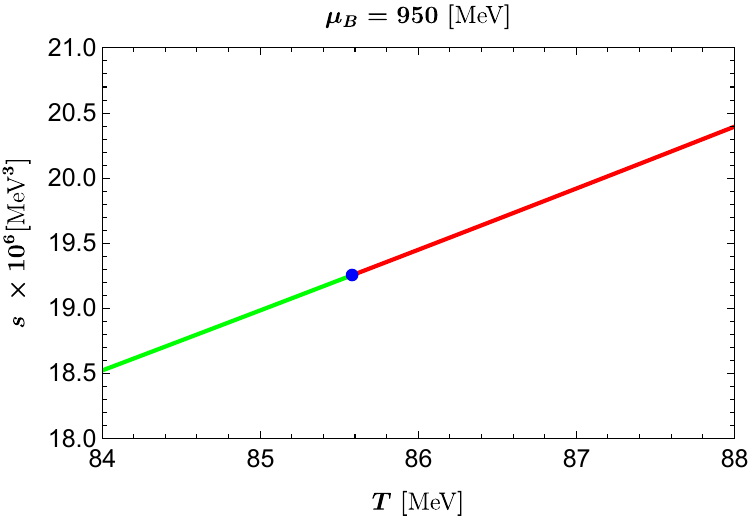}
    \caption{The entropy density with respect to temperature for different chemical potentials under the same conditions as in Fig.~\ref{fig=freeenergy}.}
    \label{fig=entropy}
\end{figure}

\begin{figure}[thbp]
    \centering
    \includegraphics[width=0.24\linewidth]{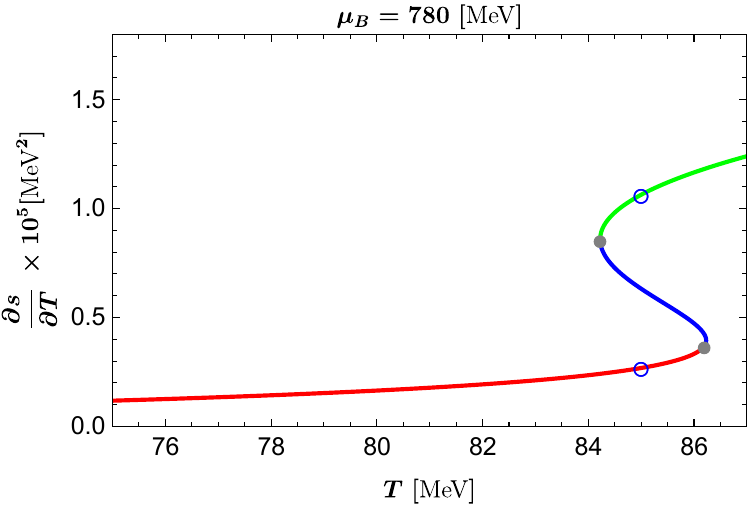}
    \includegraphics[width=0.24\linewidth]{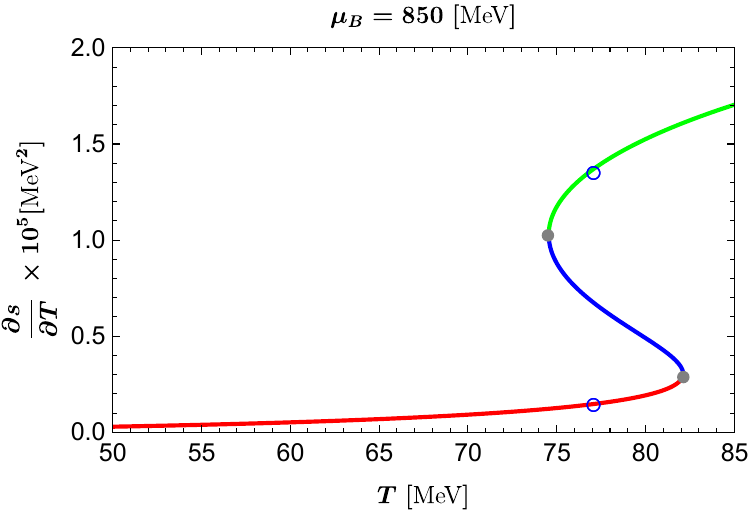}
    \includegraphics[width=0.24\linewidth]{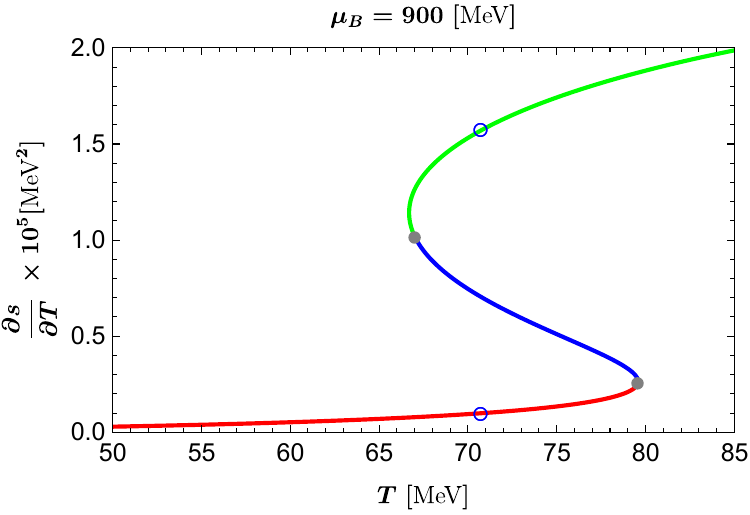}
    \includegraphics[width=0.24\linewidth]{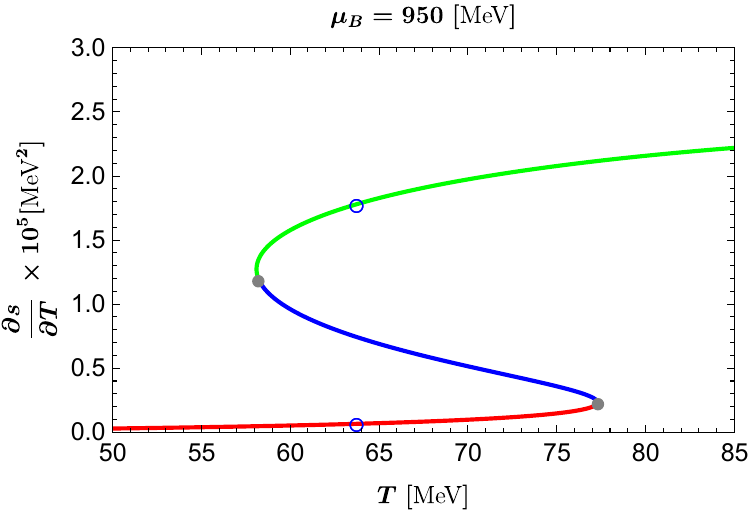}
    \includegraphics[width=0.24\linewidth]{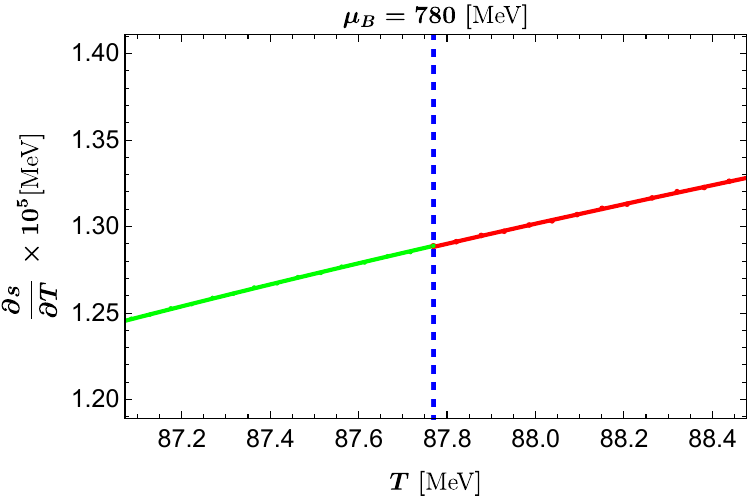}
    \includegraphics[width=0.24\linewidth]{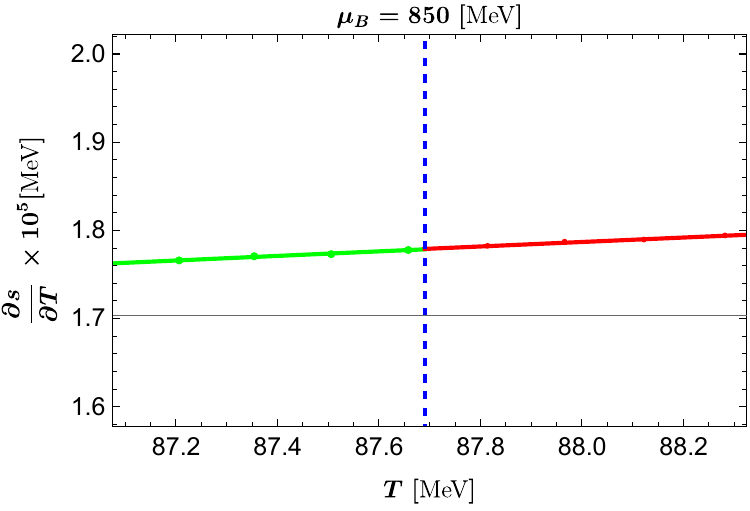}
    \includegraphics[width=0.24\linewidth]{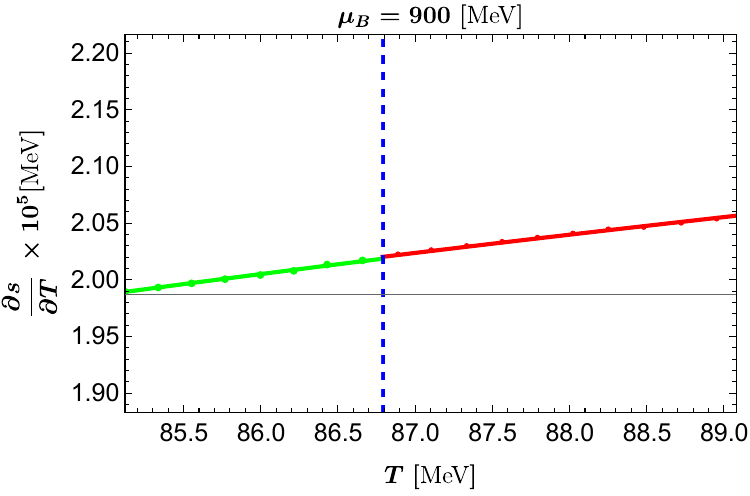}
    \includegraphics[width=0.24\linewidth]{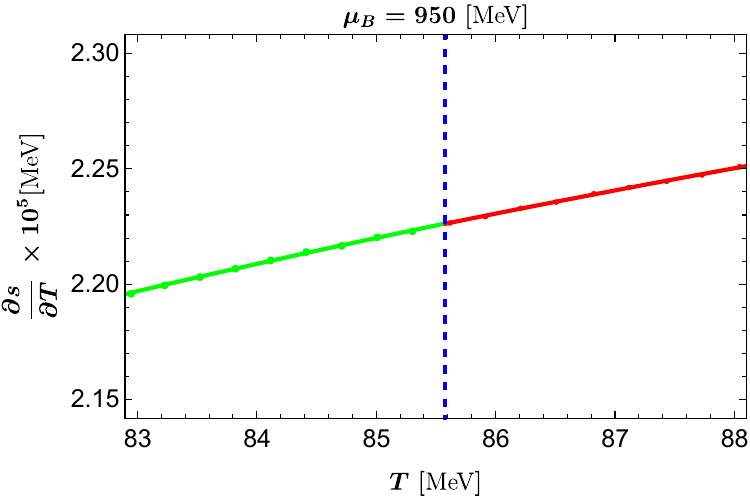}
    \caption{The first-order derivative of the entropy density with respect to temperature for different chemical potentials under the same conditions as in Fig.~\ref{fig=freeenergy}.}
    \label{fig=dentropy}
\end{figure}

\begin{figure}[thbp]
    \centering
    \includegraphics[width=0.48\linewidth]{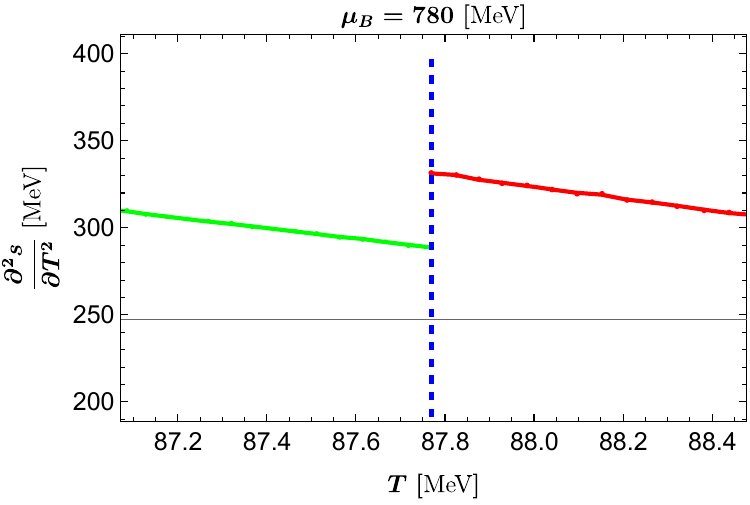}
    \includegraphics[width=0.48\linewidth]{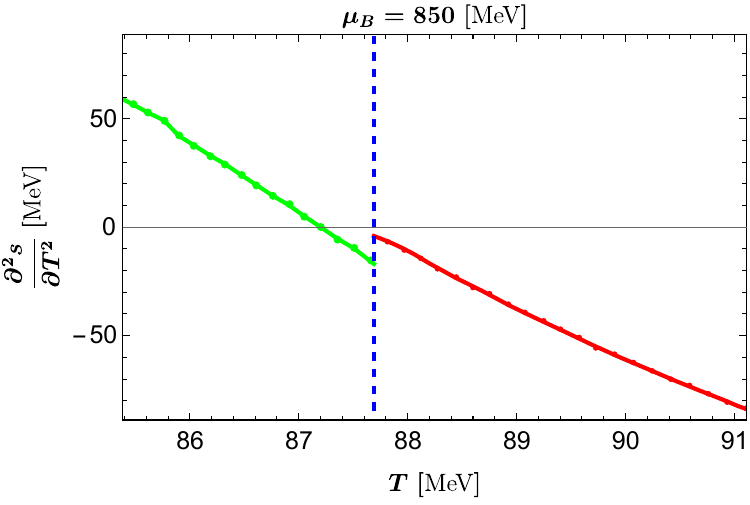}
    \includegraphics[width=0.48\linewidth]{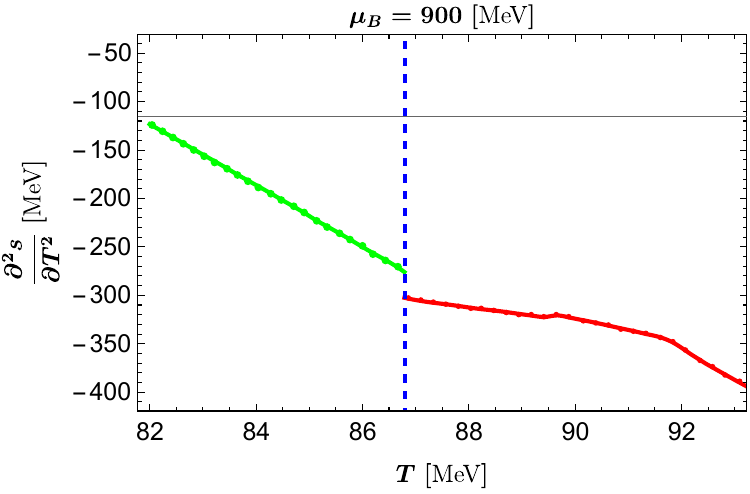}
    \includegraphics[width=0.48\linewidth]{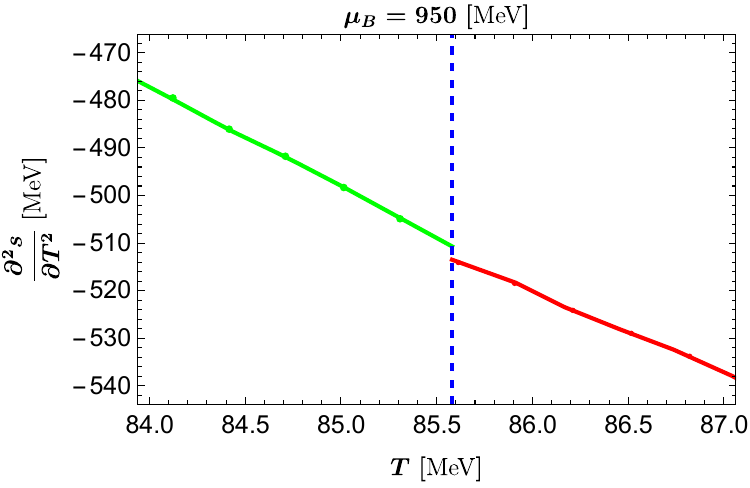}
    \caption{The second derivative of the entropy density with respect to temperature for different chemical potentials, corresponding to the bottom row of Fig.~\ref{fig=freeenergy}. 
    The discontinuities in the evolution of the entropy density clearly signal a third-order phase transition. }
    \label{fig=ddentropy}
\end{figure}

Through the holographic dictionary, thermodynamic quantities in the dual field theory at finite temperature and baryon density, including temperature $T$, entropy density $s$, baryon chemical potential $\mu_B$ and baryon charge density $\rho_B$ are derived from Eq.~\eqref{eq=thermoquan}.
For finite $\mu_B$, the internal energy density $\epsilon$ and free energy density $F$ obey
\begin{align}
	& \epsilon(s,\rho_B)=Ts-P+\mu_B\rho_B,\\
	& F(T,\mu_B)=-P(T,\mu_B)=\epsilon(s,\rho_B)-Ts-\mu_B\rho_B.
\end{align}
Consequently, at fixed chemical potential, the differential relation for the free energy density are given by
\begin{align}
	dF(T,\text{fixed}\ \mu_B)=-dP(T,\text{fixed}\ \mu_B)=sdT.
\end{align}
Applying Ehrenfest classification, one can evaluate the free energy density and its derivatives to determine the orders of the phase transition.
Therefore, in this paper, we examine the phase transitions occurring at these two boundaries by studying the free energy density and its temperature derivatives at fixed chemical potentials $\mu_B =780,850,900,950\text{MeV}$. 
The resulting free energy density $F$, its first derivative (i.e. the entropy density $s$), second- and third-order derivatives (i.e. the first- and second-order derivatives of entropy density $\frac{\partial s}{\partial T}, \frac{\partial^2 s}{\partial T^2}$) as the function of the temperature are present in Fig.~\ref{fig=freeenergy},~\ref{fig=entropy},~\ref{fig=dentropy} and~\ref{fig=ddentropy}.
The top panels of Fig.~\ref{fig=freeenergy},~\ref{fig=entropy} and~\ref{fig=dentropy} display results along the blue dashed boundary, while the bottom panels show the phase transitions in the blue solid boundary at the same $\mu_B$.

The four results along the blue dashed phase boundary clearly demonstrate characteristic features of first-order phase transitions, as illustrated in the top panels of Fig.~\ref{fig=freeenergy} and~\ref{fig=entropy}. 
In Fig.~\ref{fig=freeenergy}, the free energy density as a function of temperature displays a distinctive zigzag trajectory at the phase coexistence region.
The intersection point of these two free energy branches, marked by a blue circle, indicates the critical temperature where the actual phase transition occurs. 
Beyond this critical point, the green and red solid curves correspond to metastable subcooled and superheated states, respectively.
The emergence of the swallowtail structure in the free energy density near the phase transition temperature provides clear evidence for the first-order nature of the phase transition.

Correspondingly, Fig.~\ref{fig=entropy} reveals an S-shaped profile in the temperature dependence of entropy density.
The first-order phase transition manifests as a finite vertical jump in entropy density at the critical temperature, as indicated by the blue circle.
Furthermore, with decreasing chemical potential, both the swallowtail shape in the free energy density and the S-shaped profile in the entropy density progressively diminish, as shown by the evolution along the blue dashed phase boundary approaching the critical point.
This contraction behavior reflects the diminishing thermodynamic instability gap between the coexisting phases as the system approaches the critical point of the phase transition line.

As observed in the phase diagram in Fig.~\ref{fig=phase}, beyond this critical point the phase boundary does not terminate rather turns back, appearing in a region of relatively higher temperature.
As demonstrated in the second-row panels of Fig.~\ref{fig=freeenergy},~\ref{fig=entropy} and~\ref{fig=dentropy}, the free energy density, entropy density, and first derivative of entropy density all exhibit smooth continuous variations along the blue solid phase boundary line.
However, the behavior of the second derivative of entropy density in this domain, as revealed in Fig.~\ref{fig=ddentropy}, shows a finite discontinuity at the phase boundary.
This discontinuity identifies the blue solid phase boundary segment as representing a third-order phase transition line within our framework. 
Numerically, this third-order transition appears remarkably smooth and subtle, making it particularly challenging to detect through dual theory analysis alone across the vast parameter space.
By thoroughly analyzing the phase boundary and computing numerous hairy black hole solutions, we can precisely resolve this discontinuity and identify the nature of the phase transition occurring here.

Returning to the phase diagram in Fig.~\ref{fig=phase}, the blue curve, which delineates the boundary between the Type I and Type II hairy solutions, can now be unambiguously identified as the line of phase transitions between these two phases.
The dashed portion corresponds to the first-order phase transition line previously studied in holographic critical endpoint models, while the solid segment represents the third-order transition line emerging in our simplified EMD model.
These two transition lines terminate at a critical point $(\mu_B^{\text{crit}},T_{\text{crit}})$.
Notably distinct from the simplified EMD model, the critical points of these transition lines now coincide with the turning point of the phase boundary in the current framework.
Consequently, for a $\mu_B>\mu_B^{\text{crit}}$, the system first undergoes a sharp first-order phase transition as the temperature is raised, and then experiences a smooth and mild third-order phase transition.
The behaviors of the scalar field profiles at different temperatures, as illustrated in Fig.~\ref{fig=profile}, further validates our thermodynamic analysis.

\section{Results and further discussion}\label{sec=conclusion}

In this work, we adopt the techniques of black hole physics from~\cite{Guo:2024ymo} to analyze hairy black hole solutions and their dual phase structures in the improved holographic EMD model proposed in~\cite{Critelli:2017oub,Grefa:2021qvt}.
From the gravitational sector perspective, the rich phase structure observed in the dual field theory and its remarkable consistency with current QCD theories stem from intricate constructions of the scalar potential $V(\psi)$ and Maxwell field coupling function $f(\psi)$ through matching with lattice QCD data.
The key point is that such a constructed matter sector gives rise to different types of black hole states in the bulk, whose parameter-dependent transitions holographically encode the phase transitions in the boundary theory.

Two distinct types of hairy black hole solutions are found in this framework.
Type-I hairy black holes governed by scalar potential $V(\psi)$ dominance, where the scalar field exhibits monotonic radial decay. 
Type-II solutions arising from a nonminimal coupling to the Maxwell field, characterized by nonmonotonic scalar profiles that initially grow near the horizon, and subsequently decay after reaching a local maxima.
Boundary condition analysis demonstrates that the near-horizon growth/decay tendency of the scalar field depends critically on the sign of $\psi_h^1$.
When the potential $V(\psi)$ dominates, $\psi_h^1$ is negative, ensuring a monotonically decaying profile, whereas when the coupling function $f(\psi)$ dominates, a positive $\psi_h^1$ induces an initial growth in the scalar field.
This finding aligns with~\cite{Guo:2024ymo}'s observations in the simplified EMD model, where the scalar field influenced solely by the potential $V(\psi)$ decays monotonically and that affected solely by the coupling function $f(\psi)$ increases monotonically, approaching a constant in the far-field region.
These results conclusively validate the coexistence of two fundamentally distinct hairy black hole types.

Through the analysis of these hairy black hole solutions, we precisely map the distribution of the Type I and Type II hairy black hole states in the parameter space and identify their phase boundary, as shown in the left panel of Fig.~\ref{fig=parameter} and Fig.~\ref{fig=phase}.
This represents a challenging aspect for traditional holographic methods and highlights the advantage of the black hole physics approach.
In the $(\mu_B,T)$ plane of Fig.~\ref{fig=phase}, the phase boundary comprises two segments: The blue dashed line demarcates the overlapping phase coexistence region (green subdomain below), where first-order phase transitions occur, consistent with previous results in~\cite{Critelli:2017oub,Grefa:2021qvt}.
The blue solid line represents the non-overlapping phase boundary, where thermodynamic analysis reveals a subtle third-order phase transition.
The results can be verified in Fig.~\ref{fig=freeenergy},~\ref{fig=entropy},~\ref{fig=dentropy} and~\ref{fig=ddentropy}.
Both transition lines terminate at a critical point $(\mu_B^{\text{crit}},T_{\text{crit}})=(765.51,86.54) \text{MeV}$, which coincides exactly with the turning point of the complete phase boundary curve.
This differs from the results in the simplified EMD model~\cite{Guo:2024ymo}, where the critical point is not the turning point of the boundary line, and the overlap region of the two phases is much smaller than in the present work.
This suggests that targeted modifications to the coupling function $f(\psi)$ and potential function $V(\psi)$ could effectively engineer specific phase structures for simulating strongly coupled systems.

Our work serves as a valuable complement to existing studies of the holographic EMD model by comprehensively characterizing hairy black hole phase distributions, boundaries, and their gravitational sector interpretations. 
This complete theoretical framework is essential for a proper understanding of holographic EMD dynamics, thereby establishing a solid foundation for applying these models to phase transition studies in QCD and other strongly coupled systems.
From this perspective, a detailed analysis of how potential function parameters and the coupling coefficients affect the boundary field theory's phase structure could further enhance our understanding of the model's microscopic organization.
On the other hand, incorporating richer matter fields, such as holographic superconductors~\cite{Li:2011xja,Guo:2020sdu,Guo:2024vhq,Brihaye:2019dck} or higher-order curvature terms~\cite{Kanti:1995vq,Torii:1996yi}, may give rise to even more intricate phase structures, providing another interesting avenue for exploration.

\begin{acknowledgments}
We thank Prof. Song He for his helpful discussion.
This work is supported in part by the financial support from Brazilian agencies 
Funda\c{c}\~ao de Amparo \`a Pesquisa do Estado de S\~ao Paulo (FAPESP), 
Fundação de Amparo à Pesquisa do Estado do Rio Grande do Sul (FAPERGS),
Funda\c{c}\~ao de Amparo \`a Pesquisa do Estado do Rio de Janeiro (FAPERJ), 
Conselho Nacional de Desenvolvimento Cient\'{\i}fico e Tecnol\'ogico (CNPq), 
and Coordena\c{c}\~ao de Aperfei\c{c}oamento de Pessoal de N\'ivel Superior (CAPES).

\end{acknowledgments}

\bibliographystyle{jhep}
\bibliography{main.bib}
\end{document}